# A Novel Discovery of Negative Thermal Expansion in Rare-earth Pyrochlore through Anion Order-Disorder Transition

Yuxuan Wang, Guoqiang Lan, Jun Song[*]

*Department of Mining and Materials Engineering, McGill, University, Montreal, Quebec, H3A 0C5, Canada*

**Abstract**

In this study, we report for the first time the occurrence and investigation of the negative thermal expansion (NTE) effect in rare-earth pyrochlores. It's found that the NTE originates from the migration of oxygen anions from 48f sites to 8b sites, where one-twelfth of the original anions will gradually occupy half of the available oxygen vacancies. This initial rapid transition leads to the distortion and rotation of polyhedral units, effectively contracting the lattice and manifesting as macroscopic NTE. This transition is sensitive to external isotropic pressure, where increasing pressure delays the onset of anion transition. These study deepen our understanding of NTE in complex oxides and demonstrate the utility of deep learning potentials for exploring intricate structural behaviors.

**Graphic Abstract:**

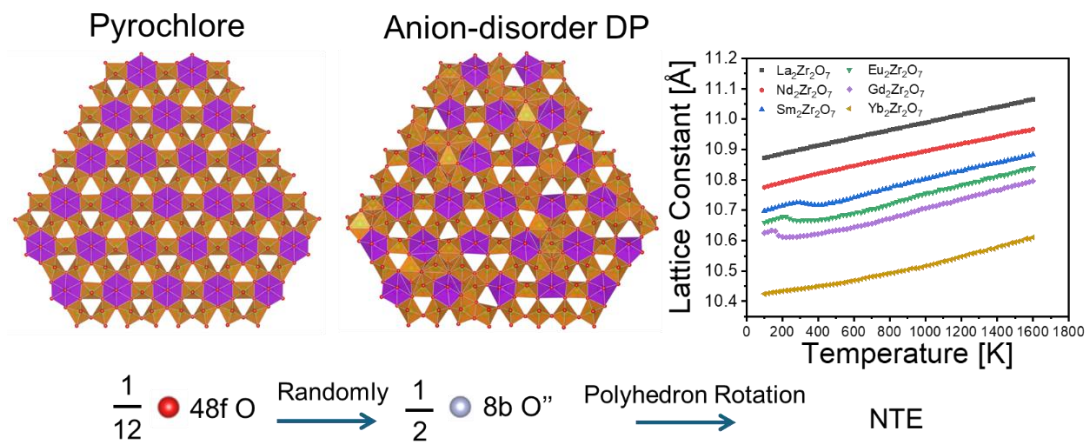

[*] Author to whom correspondence should be addressed. E-mail address: jun.song2@mcgill.ca (J. Song).

# Introduction

Pyrochlore oxides are a class of ceramic oxides of a complex crystal structure characterized by a highly symmetrical and flexible three-dimensional framework. The structural complexity allows for extensive chemical substitutions and defect engineering and render pyrochlore oxides with great functional tunability for potential applications in different fields[1-3]. The excellent ability of pyrochlore oxides to accommodate oxygen vacancies for enhancing ionic conductivity[1] make them viable candidates for solid oxide fuel cells (SOFCs) [4, 5] and oxygen-ion conductors[6, 7]. Pyrochlore oxides are also highly sought-after materials for developing novel thermal barrier coatings (TBCs) [8-13] enabling higher combustion temperatures and fuel efficiency because of their lower thermal conductivity[14, 15], superior phase stability and oxidation resistance at high temperatures. Moreover, with their tunable surface activities and electronic structures, pyrochlore oxides present potential in catalysis applications, e.g., in oxygen evolution reactions (OER)[16, 17] and $CO_2$ reduction[18, 19].

An ordered pyrochlore oxide, with the general formula $A_2B_2O_6O'$, belongs to the $Fd\bar{3}m$ space group and can be regarded as a superstructure derived from fluorite-type lattice ($Fm\bar{3}m$) with twice the lattice constant, where A-sites (16c) are often occupied by trivalent rear-earth cations and B-sites (16d) by tetravalent cations respectively, with $.\bar{3}m$ symmetry[20]. Compared to the ideal fluorite structure, one-eighth of the oxygen anions are systematically missing, leading to an ordered array of inherent oxygen vacancies (8b). For the rest of the oxygen anions, six-eighth (O) take the 48f site with $2.mm$ symmetry while one-eighth (O') occupy the 8a site with $\bar{4}3m$ symmetry. However, this ordered structure of pyrochlore oxides may be subjected to the influence of temperature effect[21] and chemical composition variation[22-24], leading to structural transition and ion disorder in these oxides. Pyrochlore oxides may undergo order-disorder transition into a defective fluorite phase at elevated temperatures, where A-site and B-site cations will be randomly distributed over the 8c cation lattice ($\bar{4}3m$ symmetry) while anions take 4a sites ($m\bar{3}m$ symmetry) with 0.875 occupancy[25, 26]. This was observed to occur in pyrochlore oxides of $Nd_2Zr_2O_7$, $Sm_2Zr_2O_7$ and $Gd_2Zr_2O_7$[21], at approximately 2300 °C, 2000 °C, and 1530 °C, respectively. The phase stability of pyrochlore is also strongly influenced by ionic radius ratio[27] between $A^{3+}$ and $B^{4+}$ cations. Empirically, when the ratio falls within the range of 1.46-1.78, a well-ordered pyrochlore phase is stabilized. When the ratio drops below 1.46, the structure typically transforms into a defect fluorite phase, whereas ratios above 1.78 tend to favor the formation of a monoclinic phase. Furthermore, substitutional doping can also trigger the order-disorder transition. For instance, Mandal et al.[28] reported that A-site Gd doping in $Nd_2Zr_2O_7$ induces a discontinuous order-disorder transition from pyrochlore to defective fluorite.

Substantial changes in physical properties often accompany above pyrochlore-defective fluorite transition. Che et al.[29] investigated the difference in thermal conductivity between the pyrochlore and defective fluorite phases using molecular dynamics (MD) simulations, revealing that the thermal conductivity of the defective fluorite phase is typically 30-50% lower than that of the pyrochlore phase of the same composition. Yang et al. [30] demonstrated that the ionic conductivity of a mixed-phase Y-doped $La_2Zr_2O_7$ system is 2-3 orders of magnitude higher than that of either the single-phase pyrochlore $La_2Zr_2O_7$ or the defective fluorite $Y_2Zr_2O_7$. Therefore, precise knowledge of the pyrochlore-defective fluorite transition is crucial in the design and application of pyrochlore oxides-based materials. Previous studies suggest that the phase transition from pyrochlore to defective fluorite be not a one-step process but involving one or more intermediate steps. Lian etc.[31] found that under ion radiation condition, the pyrochlore structure experienced an anion disordering step prior to the transition to the cation-disorder defective fluorite phase. Such anion disordering step was also confirmed in the study of A site doping process of $Dy_2Zr_2O_7$ by Li et al.[32], where the SAED patterns corresponding to anion order were observed to disappear first. Another study by Moriga et al.[33], a dynamic pyrochlore phase in $Gd_2Zr_2O_7$ annealed at 1600 °C was observed, where cations show partial disorder while the 8a anions partly move to the 8b vacancy sites. Through neutron Rietveld analysis, Catherine et al.[23] investigated the oxygen occupation of disorder induced by Zr substitution of $Y_2(Zr_yTi_{1-y})O_7$. The results indicate that with the increasement of Zr, the 48f oxygens will first transfer to 8b sites, and then cation order-disorder will occur accompanying with 8a-8b oxygen transition.

However, the exact process underlying phase transition from pyrochlore to defective fluorite remain elusive with critical mechanistic details lacking, due to limitations of experimental techniques. X-ray diffraction (XRD) lacks the resolution to identify subtle anion movement, while transmission electron microscopy (TEM) and synchrotron radiation techniques are generally not suitable for real-time observation over a wide temperature range. These limitations from the experimental side may be effectively addressed by atomistic simulations. *Ab-initio* density functional theory (DFT) calculations provide high-precision prediction of the lattice structure, energy and electronic states, while MD simulations enable real-time, large-scale investigation of the dynamic processes involved in phase transition. Nevertheless, an optimal balance between computational accuracy and efficiency has yet to be achieved by either approach. More recently, deep-learning interatomic potential models, such as Deep-Potential (DP) model[34-36], combines both the merits of DFT and MD, enabling *ab-initio* precise, large-scale and real-time simulation, thus emerging as a powerful tool for investigating phase transition mechanisms.

In this study, we successfully trained a DP model for rare-earth zirconia pyrochlore systems, comprising La, Nd, Sm, Eu, Gd, Yb, Zr, and O elements. This model enables real-time dynamic simulations, offering an effective tool to gain in-depth insight into the phase transition process. During the simulations, we observed negative thermal expansion (NTE) behavior in $Sm_2Zr_2O_7$, $Eu_2Zr_2O_7$ and $Gd_2Zr_2O_7$, indicating possible phase transitions. Further investigations revealed that, with increasing temperature, the perfect pyrochlore structure transitions to an anion-disordered subphase ($P_1$), where one-twelfth of the 48f oxygen anions gradually migrate to half of the 8b sites. This transition begins rapidly near the starting temperature and gradually completes at higher temperatures. At the onset of this transition, numerous $AO_8$ and $BO_6$ polyhedron transform into $AO_7$ and $BO_7$ configurations, leading to frame rotation and space compression, ultimately resulting in NTE. This phenomenon is sensitive to isotropic pressure; for instance, applying 6 GPa external pressure increases the starting temperature of this transition from 120 K to approximately 220 K for $Gd_2Zr_2O_7$. This anion order-disorder transition process, accompanying with negative thermal expansion phenomenon, is revealed for the first time in our study, which deepens the understanding of phase transition mechanisms and provides valuable insights for the design of pyrochlore-structured materials

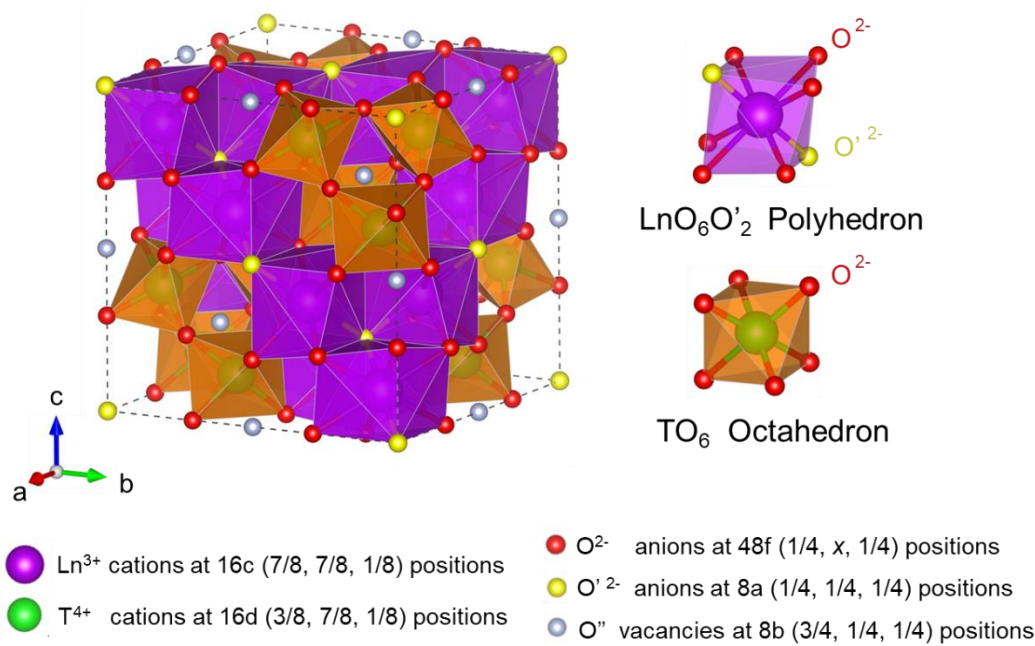

**Figure 1** Crystal structure of an ordered pyrochlore oxide with general formula $Ln_2T_2O_7$. The structure is composed of $LnO_6O'_2$ polyhedra (purple) and $TO_6$ octahedra (orange), forming a three-dimensional network. Rare-earth $Ln^{3+}$ cations (purple spheres) occupy the 16c Wyckoff positions (7/8, 7/8, 1/8), while $T^{4+}$ cations (green spheres) reside at the 16d positions (3/8, 3/8, 1/8). $O^{2-}$ anions (red spheres) are located at 48f (1/4, $x$, 1/4) positions, $O'^{2-}$ (yellow spheres) at 8a (1/4, 1/4, 1/4), and $O''$ vacancies (gray spheres) at 8b (3/4, 1/4, 1/4).

# Results

## Lattice constant and thermal expansion

Phase transitions are often manifested as changes in lattice constants. Therefore, at first, AIMD simulations was employed to calculate the lattice constants of six rare-earth pyrochlore oxides: $La_2Zr_2O_7$, $Nd_2Zr_2O_7$, $Sm_2Zr_2O_7$, $Eu_2Zr_2O_7$, $Gd_2Zr_2O_7$ and $Yb_2Zr_2O_7$. The temperature dependence of the lattice constants for $La_2Zr_2O_7$ and $Nd_2Zr_2O_7$ is presented in Fig. 1(a). At 50 K, their lattice constants are 10.84 Å and 10.77 Å, respectively. As the temperature increases, both exhibit an approximately linear increase, suggesting the absence of phase transitions. By 1800 K, their lattice constants reach 11.10 Å and 11.00 Å, respectively. And for the other four compositions, their curves exhibit distinct trends, as shown in Fig. 2(b). While the lattice constants of all four compounds generally increase with temperature, their temperature dependence deviates from linearity. Specifically, $Sm_2Zr_2O_7$ and $Eu_2Zr_2O_7$ exhibit a rapid increase at low temperatures, followed by a slower increase at higher temperatures. $Gd_2Zr_2O_7$ demonstrates a rapid increase within the temperature ranges of 50-300 K, 600-1200 K, and above 1500 K, while showing a relatively slower increase in the intermediate ranges of 300-600 K and 1200-1500 K. In the case of $Yb_2Zr_2O_7$, the lattice constant increases gradually, with a continuously increasing rate.

Despite these variations in lattice expansion trends observed from AIMD calculations, the limited number of temperature points and limited relaxation times restrict the ability to comprehensively investigate potential phase transitions. To address these limitations, based on these AI-MD results, deep-learning potential through DP model[37, 38] were successfully trained, enabling continuous dynamic simulations with larger supercells and extended relaxation times. The results are shown in Fig. 2(c). In temperature range from 100K to 1600K, we can see that for $La_2Zr_2O_7$ and $Nd_2Zr_2O_7$, the lattice constant curves are almost perfectly linear and align well with AIMD results, suggesting no phase structure change. However, for $Sm_2Zr_2O_7$, $Eu_2Zr_2O_7$ and $Gd_2Zr_2O_7$, the lattice constants show prominent valleys at low temperatures, indicating possible phase transitions. The temperature of these valleys shifts downward with increasing atomic numbers of Sm, Eu, and Gd, from approximately 400 K to 200 K. In the case of $Yb_2Zr_2O_7$, although no clear valley is observed, its lattice constant curve deviates from the linear trends seen in $La_2Zr_2O_7$ and $Nd_2Zr_2O_7$.

To further investigate the change of lattice constant, coefficients of thermal expansion (CTE) were plotted and shown in Fig. 2 (d). For $La_2Zr_2O_7$ and $Nd_2Zr_2O_7$, the CTE initially decreases slightly and then stabilizes around $11 \times 10^6$ K$^{-1}$ at higher temperatures. In contrast, $Sm_2Zr_2O_7$, $Eu_2Zr_2O_7$ and $Gd_2Zr_2O_7$ show distinctly different CTE behaviors. Their CTE curves decrease initially, reaching a minimum negative value, then rise back to positive values and stabilize at higher temperatures. At 1500 K, the CTEs for $Sm_2Zr_2O_7$, $Eu_2Zr_2O_7$ and $Gd_2Zr_2O_7$ are $12.5 \times 10^6$ K$^{-1}$, $13.0 \times 10^6$ K$^{-1}$ and $14.2 \times 10^6$ K$^{-1}$, respectively. For $Yb_2Zr_2O_7$, the CTE starts at around $7.5 \times 10^6$ K$^{-1}$ at low temperature, increases steadily, and stabilizes at high temperatures, reaching $15.1 \times 10^6$ K$^{-1}$ at 1500 K. These observations suggest that a specific mechanism may contribute to the negative thermal expansion behavior observed in $Sm_2Zr_2O_7$, $Eu_2Zr_2O_7$ and $Gd_2Zr_2O_7$.

In fact, many crystal structure of complex oxides have been found with negative thermal expansion (NTE)[39]. Traditional mechanism includes anisotropic NTE such as $\beta$-eucryotite[40] and cordierite[41]; flexible network contraction like $ZrW_2O_8$[42] and $ZrV_2O_7$[43]. And phase transition type like ferroelectric transition[44] ($PbTiO_3$-$BiFeO_3$), charge-transfer transition[45] ($SrCu_3Fe_4O_{12}$) and metal-insulator transition[46] ($CaRuO_4$) mechanisms were found later. To identify the mechanism behind pyrochlore structure, we need further investigation.

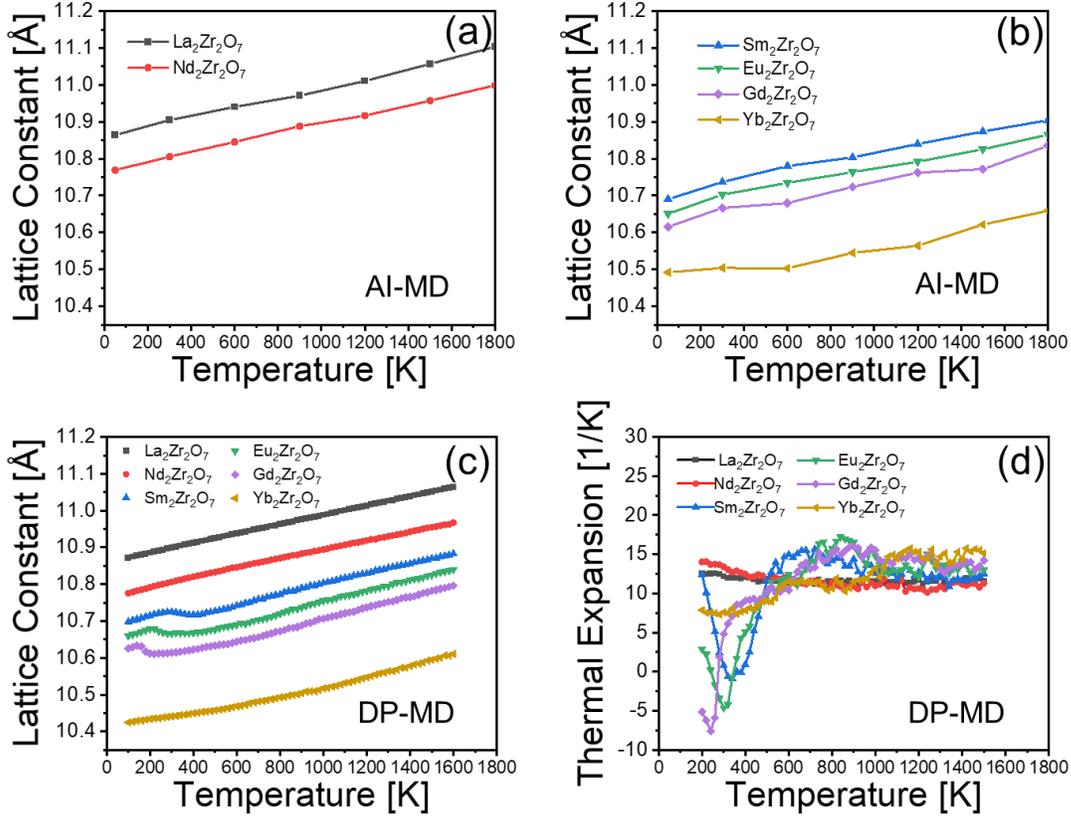

**Figure 2** Temperature dependence of the lattice constant and thermal expansion behavior for various $Ln_2Zr_2O_7$ pyrochlores. (a-b) Lattice constants obtained from AI-MD simulations. (c) Lattice constants derived from deep DP-MD simulations. (d) Corresponding thermal expansion coefficients calculated based on the temperature derivatives of the lattice constants in (c).

## XRD simulations

To investigate the potential phase transition, $Gd_2Zr_2O_7$ was selected as representative. Its XRD simulations results at various temperatures are presented in Fig. 3. In experimental XRD tests, peaks corresponding to the (331) and (511) planes are typically used to differentiate between pyrochlore and defective fluorite structures[22]. The disappearance of these two peaks indicates that the ordered arrangement of A and B site cations is disrupted, making them indistinguishable. However, for oxygen anions, their scattering factors are overshadowed by those of heavier atoms with significantly larger atomic numbers, rendering them usually unresolvable in conventional XRD experiments. For $Gd_2Zr_2O_7$, the scattering intensity of main peaks, such as (222), can reach up to 80 a.u., whereas peaks associated with anions ordering, such as (220) and (422), have intensities of only about 0.3 a.u. Under experimental conditions, these minor peaks are often obscured by external noise. Our simulation results, however, clearly capture the changes in all peaks.

At 100 K and 120 K, all peaks align with the standard pyrochlore pattern, including the (331) and (511) peaks, indicating that A and B site cations remain in an ordered state. Peaks associated with anion ordering, such as (220), (422), and (822), are also distinct. However, as the temperature increases to 140 K, these anion-order-related peaks disappear, and significant background noise emerges at their corresponding positions. This temperature range corresponds precisely to the turning point in the lattice constant curve for $Gd_2Zr_2O_7$ shown in Fig. 2(b). The disappearance of these peaks aligns with observations in Lian's research[31] on radiation effects and Li's study[32] on doping conditions, suggesting that temperature alone can induce a similar anion order-disorder transition under ideal conditions.

As the temperature further increases to 160 K, and even as high as 760 K, no significant changes in the peaks are observed. This indicates that the phase transition begins at 140 K and remains stable at higher temperatures.

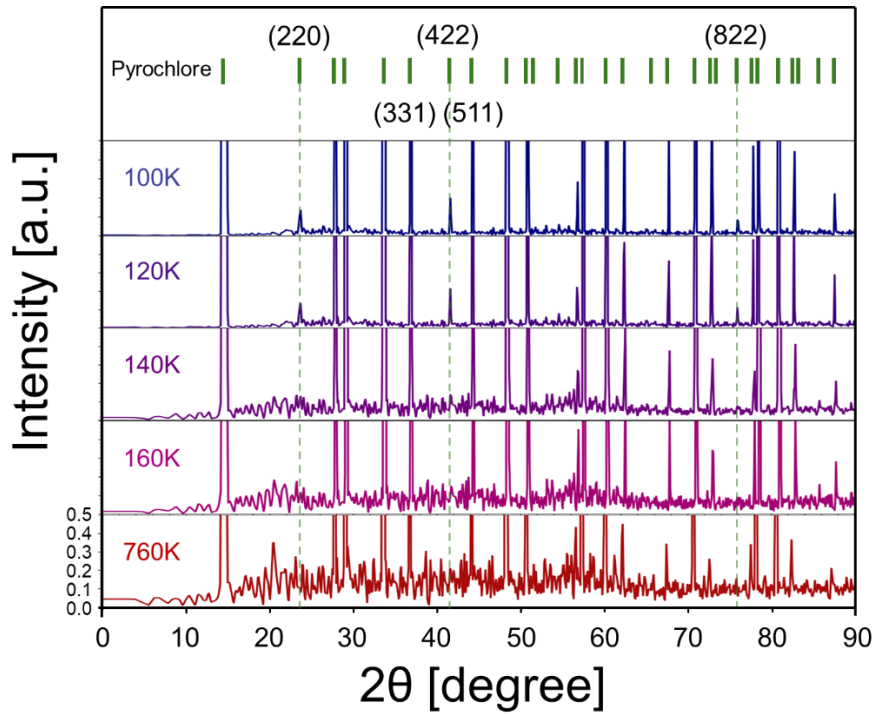

**Figure 3** Simulated X-ray diffraction patterns of $Gd_2Zr_2O_7$ at various temperatures ranging from 100 K to 760 K. Characteristic reflections of ordered pyrochlore structure are indicated by vertical green bars, with representative planes such as (220), (331), (422), (511), and (822) labeled.

Investigation on anion transition

To further investigate the transition mechanism in detail, we conducted an in-depth study on the configurational changes of $GdO_8$ and $ZrO_6$ complex polyhedron. By analyzing the relative vibrational position distributions of specific $GdO_8$ and $ZrO_6$ complexes at various temperatures (800 frames for every temperature), we aimed to observe potential anion position transfers. The results are presented in Fig. 4 and Fig. 5.

In Fig. 4, the results for two projection directions, [11-1] and [-121], are shown in the first row (Fig. 4(b-d)) and the second row (Fig. 4(e-g)), respectively, with corresponding schematic of $GdO_8$ complex structure illustrated in Fig. 4(a). To distinguish ion types, dashed circles are used to mark key ions, with colors corresponding to those in the schematic. For example, red represents 48f $O^{2-}$, yellow represents 8a $O'^{2-}$, purple represents $Gd^{3+}$, and gray represents 8b vacancy positions. At 120 K, as shown in Figs. 4(b) and 4(e), all ions are distributed near their ideal positions in the pyrochlore structure, marked by red dots. However, their vibrational shapes vary. For $Gd^{3+}$, the [11-1] projection exhibits a circular distribution, while the [-121] projection shows an elliptical shape, as highlighted in the zoomed-in images at the bottom right. This suggests that the vibration of $Gd^{3+}$ primarily occurs in the (11-1) plane. Meanwhile, the vibration shape of the surrounding 48f $O^{2-}$ ions indicates that their vibrations are mainly directed toward the central $Gd^{3+}$.

When the temperature increases to 140 K, Fig. 4(c) reveals that the lower-right $O^{2-}$ moves away from center $Gd^{3+}$ along the Gd-O direction. Fig. 4(f) further confirms that this 48f $O^{2-}$ shifts to the 8b vacancy position. At 160 K, this transition is fully completed, and all ions stabilize around their new positions. Besides the lower-right 48f $O^{2-}$, the other 48f $O^{2-}$ anions undergo minor displacements, and their vibration shapes also change. For $Gd^{3+}$, its equilibrium position shifts slightly in the opposite direction to the migrated $O^{2-}$. This suggests that the stable $GdO_8$ configuration transitions into $GdO_7$, leading to a new equilibrium center. The positions of the upper and lower 8a $O'^{2-}$ ions show small changes to accommodate the new $GdO_7$ configuration.

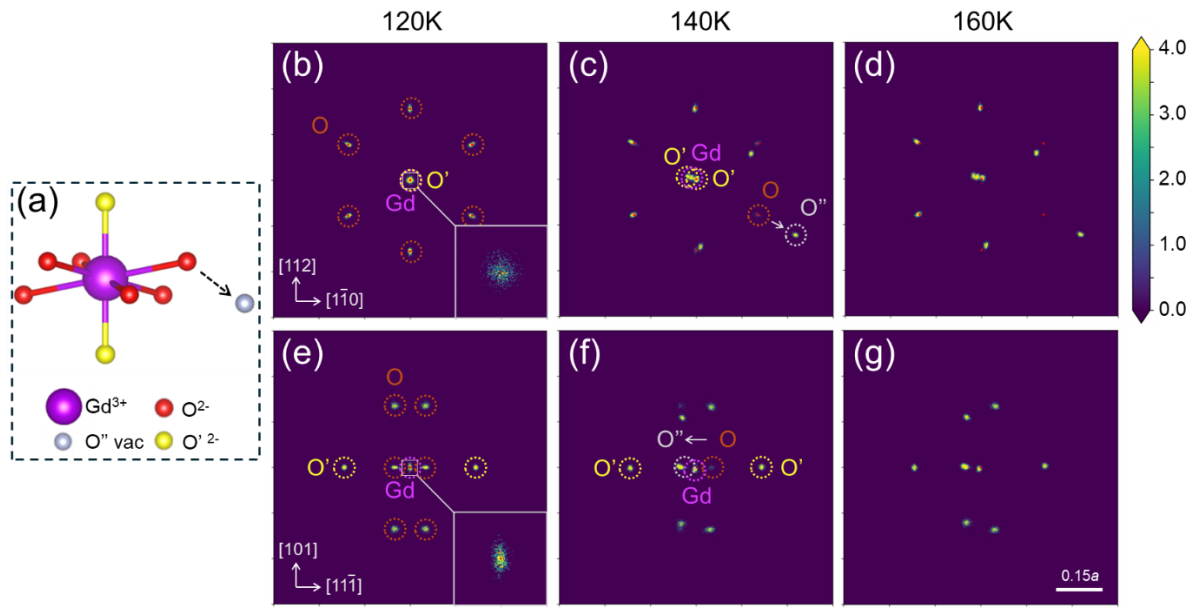

**Figure 4** Vibrational distributions of the $GdO_8$ polyhedral complex at different temperatures. (a) Schematic of the $GdO_8$ coordination environment, with $Gd^{3+}$ (purple), $O^{2-}$ (red), $O'^{2-}$ (yellow) and O'' vacancy (gray). (b-g) Vibrational probability density maps projected along the $[11\bar{1}]$ direction (b-d) and $[\bar{1}21]$ direction (e-g) at 120 K, 140 K, and 160 K respectively.

For $ZrO_6$ octahedra, the results are presented in Fig. 5. A schematic representation of the coordinated octahedron is shown in Fig. 5(a), where red spheres represent 48f $O^{2-}$, the green sphere denotes 16d $Zr^{4+}$, and the gray sphere marks the 8b vacancy position. Projection images along the [1-11] and [121] directions at 180 K, 200 K, and 220 K are illustrated in Figs. 5(b-d) and Figs. 5(e-g), respectively. Dashed circles, color-coded as in Fig. 5(a), indicate ion types.

At 180 K (c.f. Fig. 5(b) and (e)), the positions of the anions exhibit minimal deviation from the ideal pyrochlore structure, as marked by red dots. However, as the temperature increases to 200 K, notable positional changes are observed. In Fig. 5(c), the distribution of the upper $O^{2-}$ anions becomes dimmer, indicating a redistribution towards the center of the $Zr^{4+}$ cation. This behavior is confirmed in the projection along the [121] direction (c.f. Fig. 5(f)), where the anion is observed to jump into a nearby 8b vacancy. Additionally, for other 48f $O^{2-}$ anions, secondary distributions emerge near their original positions, reflecting discrete positional transitions. This discontinuous redistribution suggests that the process involves a "jump" of the 48f $O^{2-}$ anions to new positions, rather than a gradual migration. At 220 K (c.f. Fig. 5(d) and (g)), the ions stabilize around their new equilibrium positions.

In the pyrochlore structure, $GdO_8$ and $ZrO_6$ polyhedron are intrinsically linked through shared bonds, with each 48f $O^{2-}$ anion coordinated by two $Gd^{3+}$ and two $Zr^{4+}$ cations. Consequently, the configurational rearrangement of a single $GdO_8$ polyhedron inevitably affects the neighboring $ZrO_8$ octahedra. This can be confirmed by schematic both in AI-MD and DP-MD simulation of Supplementary materials (c.f. Fig. S1 and S2), where both the $GdO_7$ and $ZrO_7$ polyhedra was observed.

In summary, the configurational changes in $GdO_8$ and $ZrO_8$ polyhedron reveal that part of the 48f $O^{2-}$ anions undergo positional transitions to occupy adjacent 8b vacancies. This process is not confined to a specific temperature but occurs gradually over a broad range from approximately 120 K to 220 K. To further elucidate this phenomenon, the average displacement of oxygen anions will be counted and analyzed in histogram.

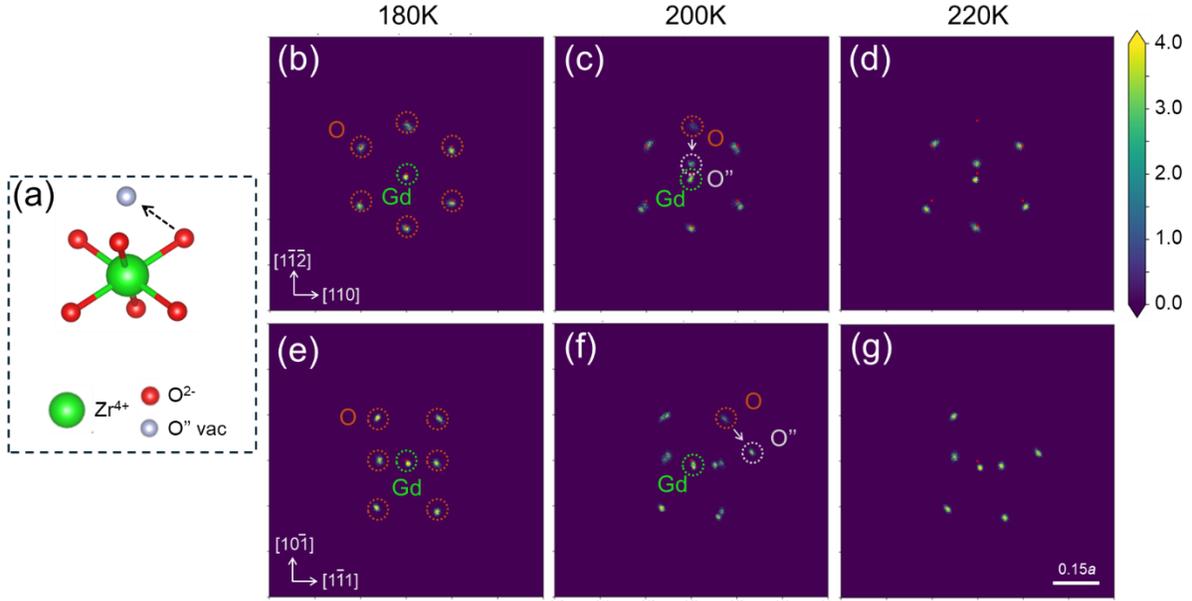

**Figure 5** Vibrational distributions of the ZrO$_6$ polyhedral complex at different temperatures. (a) Schematic of the ZrO$_6$ coordination environment, with Zr$^{4+}$ (green), O$^{2-}$ and O'' vacancy (gray). (b-g) Vibrational probability density maps projected along the [1$\bar{1}$1] direction (b-d) and [121] direction (e-g) at 180 K, 200 K, and 220 K respectively.

## Quantitative analysis

Previous qualitative analysis has shown that, in Gd$_2$Zr$_2$O$_2$, a fraction of 48f O$^{2-}$ anions transitions to fill the nearest 8b vacancies with elevated temperature. In this section, we would quantitatively analyze the position displacements of oxygen anions during the transition. This analysis is based on average fractional coordinates of 800 simulation frames for every temperature point, with the results presented in Fig. 6(a).

At 120 K, the average positions of all oxygen anions align perfectly with the pyrochlore structure, resulting in a distribution concentrated at zero displacement. However, at 140 K, two distinct distributions emerge: one concentrated at a small displacement range and the other, with a smaller count, at a larger displacement. This observation aligns with earlier findings, where the larger displacement corresponds to oxygen anions that have undergone the 48f-to-8b transition. In Fig. 6(a), the scale of x-axis represents relative displacement normalized by the lattice constant a. Therefore, assuming a lattice constant of 10.6 Å for Gd$_2$Zr$_2$O$_2$, the small displacement range corresponds to values below 0.636 Å (10.6×0.06), while the displacement of most 8b O''$^{2-}$ anions falls between 1.59 Å (10.6×0.15) and 1.91 Å (10.6×0.18). Theoretically, the distance between the 48f and 8b sites should be 2.26 Å, which is larger than the actual values. This indicates that the transition involves a smaller displacement, likely accompanied by the compression and rotation of the surrounding polyhedral complex.

To quantify the phase transition, we calculated the ratio of migrated 48f $O^{2-}$ anions. At 140 K, this ratio was 3.57%. As the temperature increases, the rate of migration slows, indicating that the transition proceeds rapidly at lower temperatures and stabilizes at higher temperatures. The transition's endpoint can be inferred from the lattice constant curve shown in Fig. 6(b). By fitting the high-temperature stable region with a straight line, the tangent point indicates the completion of the transition, with an end temperature of approximately 760 K. At this temperature, the calculated ratio of migrated anions is 7.21%, which is close to 1/14. Given that 48f $O^{2-}$ anions constitute 6/7 of all oxygen anions, it can be concluded that approximately 1/12 of the 48f $O^{2-}$ anions eventually migrate to 8b vacancy sites, which means half of the available 8b vacancies will be filled. Considering that each $Gd^{3+}$ and $Zr^{4+}$ cation is coordinated by six 48f $O^{2-}$ anions, and each 48f $O^{2-}$ anion forms bonds with four polyhedron, half of the cations at both the A-site ($Gd^{3+}$) and B-site ($Zr^{4+}$) will undergo a configurational change, corresponding to discrete displacement distribution shown in supplementary materials (c.f. Fig. S3).

The same approach can be applied to analyze other pyrochlore systems in Fid. 6(b), such as $La_2Zr_2O_7$ and $Yb_2Zr_2O_2$. In $La_2Zr_2O_7$, the lattice constant remains a perfect straight line over the entire temperature range, indicating that no phase transition occurs. In contrast, for $Yb_2Zr_2O_2$, the transition appears to start below 100 K and end around 1100 K. These conclusions are further supported by the configurational changes shown in Fig. S2 (supplementary materials) for $La_2Zr_2O_7$, $Gd_2Zr_2O_7$ and $Yb_2Zr_2O_2$ at 100 K and 220 K respectively.

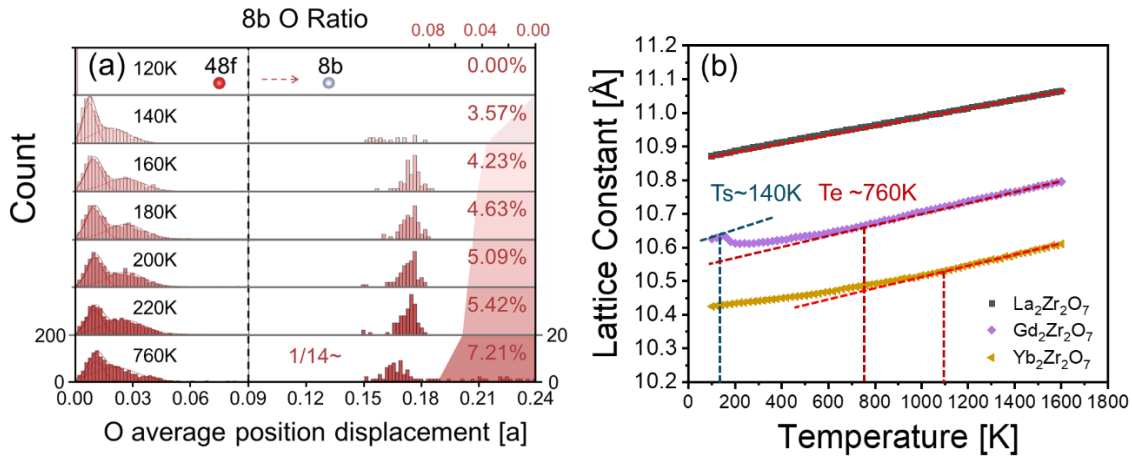

**Figure 6** (a) Distribution of oxygen displacement in $Gd_2Zr_2O_7$ at various temperature. The increasing fraction of 8b-site occupation (left red area) with temperature indicates a progressive anionic disorder, reaching approximately 1/14 occupancy at 760 K. (b) Temperature-dependent lattice constants of La$_2$Zr$_2$O$_7$, Gd$_2$Zr$_2$O$_7$, and Yb$_2$Zr$_2$O$_7$ obtained from simulations. The onset and end temperature of Gd$_2$Zr$_2$O$_7$ disorder transition are approximated as 140K and 760K referring dashed straight lines.

## Transition mechanism

From the above analysis and discussion, we have outlined the detailed transition process. This process is visualized in Fig. 7(a) and (b). In Fig. 7(a), the perfect pyrochlore structure is projected along the [111] direction, with 8b vacancies marked as gray spheres distributed across the upper and lower layers. All nearest 48f O$^{2-}$ anions have the potential to migrate and fill these vacancies. In Fig. 7(b), the configuration of the anion-disordered pyrochlore structure is depicted. In this phase, the bottom GdO$_8$ polyhedron transitions to a GdO$_7$ configuration, while the top ZrO$_6$ polyhedron converts into ZrO$_7$. This transition leads to compression and rotation of the polyhedral framework, which consequentially manifests as a macroscopic effect of negative thermal expansion.

From the perspective of energy evolution, a stable phase is generally favored when the structure exhibits lower energy. To unveil the underlying mechanism, the energies of defective pyrochlore (DP) structures, characterized by single oxygen 48f-8b transition, were calculated and compared with those of ordered pyrochlore (P) phase. The results for Gd$_2$Zr$_2$O$_7$, Sm$_2$Zr$_2$O$_7$ and Nd$_2$Zr$_2$O$_7$ under varying lattice strain are presented in Fig. 7(d-f), respectively. With increasing lattice strain, the energies of both the DP and P phases rise exponentially across all compositions. However, the relative energy relationship between DP and P differs among the three kind of pyrochlores. For Gd$_2$Zr$_2$O$_7$, P phase exhibits consistently higher energy than the DP phase when the lattice strain is below 0.03. As the strain increases, the energy gap between the two phases narrows. In the case of Sm$_2$Zr$_2$O$_7$, the P phase initially possesses slightly lower energy than the DP phase at zero strain. Upon further strain, the energy of the P phase increases and gradually approaches that of the DP phase. Conversely, for Nd$_2$Zr$_2$O$_7$, the DP phase maintains higher energy than the P phase throughout the entire strain range. These results explain the absence of negative thermal expansion in Nd$_2$Zr$_2$O$_7$ and the phase transition in Gd$_2$Zr$_2$O$_7$ and Sm$_2$Zr$_2$O$_7$, as observed in Fig. 2(c), from an energetic perspective.

However, these findings do not account for the differences in transition temperatures and rates among the compositions. To further investigate this, the energy evolution of 48f-to-8b transition was examined using interpolated static calculations between P and DP structure. The corresponding results for Gd$_2$Zr$_2$O$_7$, Sm$_2$Zr$_2$O$_7$ and Nd$_2$Zr$_2$O$_7$ are shown in Fig. 7(g-i), respectively.

An energy barrier associated with the 48f-to-8b oxygen migration is observed in all three cases, and this barrier varies with both lattice strain and composition. For $Gd_2Zr_2O_7$ (Fig. 7(g)), the energy barrier decreases from 0.49 eV to 0.25 eV with increasing strain. Similarly, for $Sm_2Zr_2O_7$, the barrier reduces from 0.58 eV to 0.40 eV. And for $Nd_2Zr_2O_7$, the barrier decreases from 0.70 eV to 0.54 eV. At the same level of lattice strain, $Gd_2Zr_2O_7$ exhibits a lower energy barrier compared to $Sm_2Zr_2O_7$, which explains its lower transition onset temperature and faster transition rate. Furthermore, once the transition occurs, polyhedral rotations lead to a volume contraction, which further reduces the system's total energy.

Based on the mechanism described above, external pressure is expected to compress the pyrochlore lattice and elevate the energy barrier for anion migration, thereby delaying the onset of the 48f-to-8b transition. This phenomenon is then confirmed by PD-MD simulations conducted under varying hydrostatic pressures. As shown in Fig. 7(c), the onset temperature of the transition increases from approximately 120 K to 220 K as the applied hydrostatic pressure rises from 0 to 6 GPa.

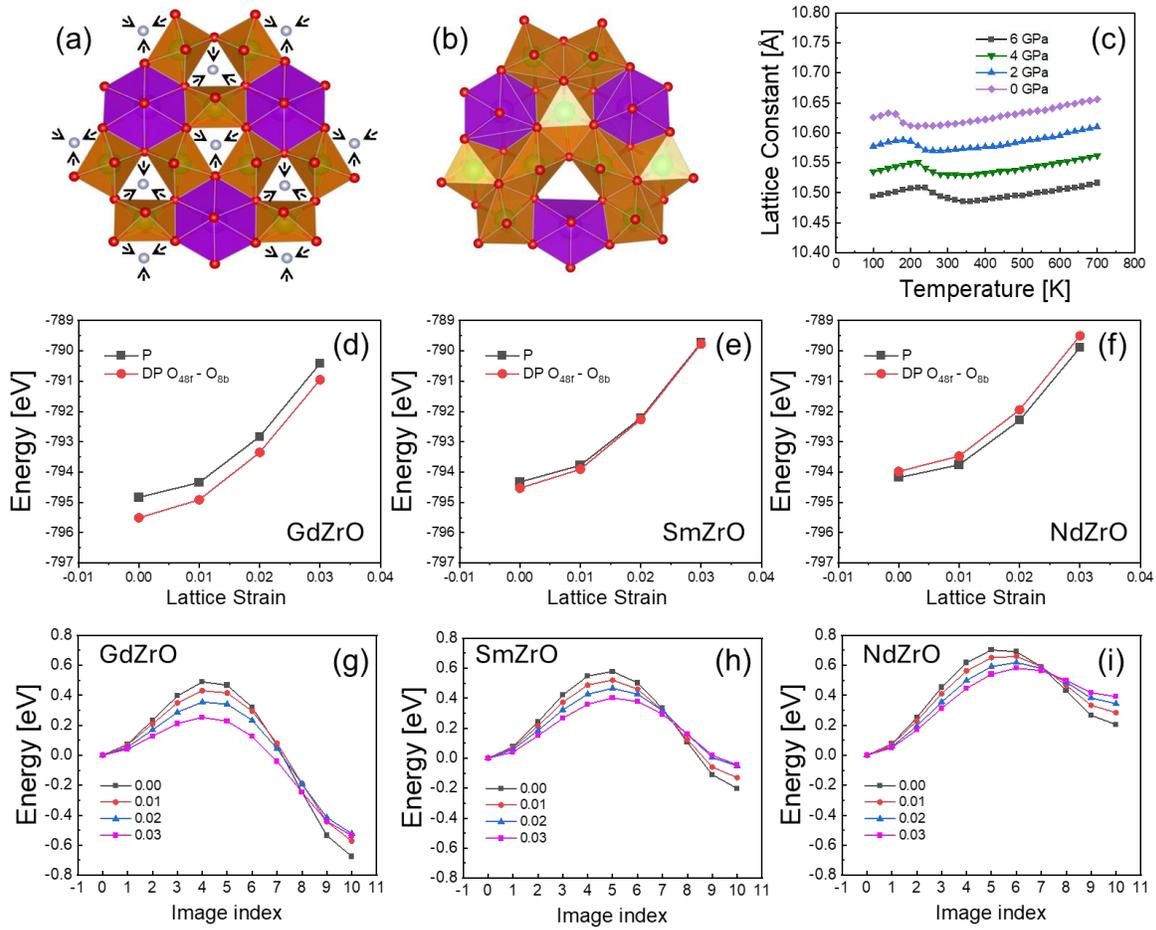

**Figure 7** (a-b) Schematics of polyhedra rotation and crystal compression in $Gd_2Zr_2O_7$ crystal. (c) Temperature dependence of lattice constant in $Gd_2Zr_2O_7$ structure under different hydrostatic pressures. (d-f) Lattice strain dependence of the potential energy in ordered pyrochlore (black square) and defective (red circle) of $Gd_2Zr_2O_7$, $Sm_2Zr_2O_7$, and $Nd_2Zr_2O_7$ respectively. In the defective configuration, single 48f $O^{2-}$ transitions to the nearest 8b site. (g-i) Corresponding energy profiles along the interpolated configurations from pyrochlore to defective cell.

## Discussion

In this study, anion order-disorder transition process in rare-earth pyrochlore oxides is systematically studied for the first time. It is concluded that with the increase of temperature, one-twelfth of the 48f oxygen anions will gradually migrate to half the sites of the 8b vacancies, leading to configurational changes and the rotation of $AO_8$ and $BO_6$ polyhedron, which macroscopically result in a negative thermal expansion effect.

This study represents the first observation and elucidation of order-disorder transition process in pyrochlore structures. These findings provide valuable insights into the interplay between lattice dynamics, volume changes, and configurational transitions, offering significant implications for understanding phase stability and functional properties in related materials. For example, previous theoretical studies[47-50] on the thermophysical properties of pyrochlores were based on idealized perfect structures and neglected the effects of anion order-disorder transitions. Our research offers a more accurate depiction of the defective pyrochlore phase, bridging this knowledge gap. In the context of solid oxide fuel cells[1, 4, 5], the 48f-to-8b anion transition pathway offers new perspectives for designing enhanced ion conduction properties. Additionally, the observed NTE effect highlights potential applications in thermal management materials[51-55].

However, the potential model trained in this study is only applicable below 1500 K, leaving the full transition from the ordered pyrochlore structure to the defective fluorite structure, typically occurring at higher temperatures (~1800 K), unexplored. Future work should focus on developing potential models for higher temperature to comprehensively examine all possible transition mechanisms in pyrochlore system.

## Methods

### *Ab-initio* calculations

The training dataset for the Deep Potential model was generated using ab-initio molecular dynamics (AIMD) simulations performed with the Vienna Ab initio Simulation Package (VASP)[56]. Pyrochlore supercells of five compositions - $La_2Zr_2O_7$, $Nd_2Zr_2O_7$, $Sm_2Zr_2O_7$, $Eu_2Zr_2O_7$, $Gd_2Zr_2O_7$ and $Yb_2Zr_2O_7$, were constructed by expanding the primitive unit cell into a 2 × 2 × 2 supercell (176 atoms). For each composition, AIMD simulations were conducted in the NPT ensemble at 50 K, 300 K, 600 K, 900 K, 1200 K, 1500 K, and 1800 K for 1 ns with timestep of 1 ps.

Density functional theory (DFT) calculations employed the generalized gradient approximation (GGA) with the Perdew–Burke–Ernzerhof (PBE) exchange-correlation functional [57]. The electron - core interactions were described using the projector augmented wave (PAW) method with the frozen-core approximation[58]. The valence electron configurations of the constituent elements were defined as follows: La - $5s^25p^66s^25d^1$, Nd - $5s^25p^64f^16s^2$ ($4f^3$ frozen in the core), Sm - $5s^25p^64f^16s^2$ ($4f^5$ frozen in the core), Eu - $5s^25p^64f^16s^2$ ($4f^6$ frozen in the core) , Gd - $4f^65d^16s^2$ ($4f^1$ frozen in the core), Yb - $5s^25p^64f^16s^2$ ($4f^{13}$ frozen in the core), Zr - $4s^24p^65s^24d^2$ and O - $2s^22p^4$.

The DFT calculations utilized a plane-wave energy cutoff of 520 eV and a Γ-centered k-point mesh of 1 × 1 × 1 for Brillouin-zone integration. And the electronic convergence criterion was set to an energy threshold of $5 \times 10^{-5}$ eV/atom.

**Deep-potential training**

The Deep Potential model for rare-earth pyrochlore was trained using the *DeepMD-kit* [37, 38] package, incorporating the DPA-1 descriptor - a novel attention mechanism highly effective for capturing the conformational and chemical spaces of atomic systems. The embedding network consists of three layers with 25, 50, and 100 nodes in each layer, respectively. The fitting network includes three fully connected layers, each with 200 nodes. The cutoff distances were set to be (9.0 Å, 11.0 Å).

For training, 80% of the AIMD data were used, while 10% and 10% were allocated for validation and testing, respectively. The testing results for various pyrochlore compositions are presented in Fig. S4 (supplementary materials), demonstrating that both the root mean square error (RMSE) of energy and force achieved highly accurate levels.

**Molecular dynamics**

To enable real-time Deep Potential prediction, the DeePMD-kit plugin was integrated into the GPU version of LAMMPS[59]. The supercells for molecular dynamics (MD) simulations were constructed by expanding the conventional unit cell into a 3 × 3 × 3 supercell, consisting of 2376 atoms. Simulations were performed in the NPT ensemble at various temperatures with a timestep of 1 ps.

For lattice constant calculations, temperature increments followed a stepwise process. The starting and ending temperatures were set to 100 K and 1600 K, respectively. Initially, a 5 ns simulation was performed at 100 K to allow lattice relaxation and achieve equilibrium. Subsequently, the temperature increased by 20 K over 2 ns to reach 120 K, followed by an 8 ns simulation at 120 K to maintain equilibrium. The last 2 ns at 120 K was used to calculate the lattice constant. This process was then repeated incrementally for the entire temperature range. For detailed configuration analysis, at each temperature, a 40 ns simulation was conducted. Atomic positions were recorded every 50 ps, resulting in a total of 800 frames for analysis.

## Data availability

## Acknowledgements

The author acknowledges financial support by National Science and Engineering Research Council of Canada (Grant #: NSERC RGPIN-2023-03628), and the Digital Research Alliance of Canada for providing computing resources.

## Author contributions

**Yuxuan Wang:** Conceptualization and Design; Data Generation and Experimentation; Data Analysis; Visualization and Figures; Writing and Editing; Interpretation of Results. **Guoqiang Lan:** Theoretical or Computational Support. **Jun Song:** Supervision, Funding acquisition.

## Competing interests

The authors declare no competing interests.